\documentclass[aps,prd,preprintnumbers,nofootinbib,twocolumn]{revtex4}
\usepackage{bm}
\usepackage{latexsym}
\usepackage{dcolumn}
\usepackage{amsmath,amsfonts,amssymb}
\usepackage{graphicx,epsfig}
\usepackage{psfrag}
\usepackage{amsthm}

\def\be {\begin{equation}}
\def\ee {\end{equation}}
\def\bea {\begin{eqnarray}}
\def\eea {\end{eqnarray}}
\def\bc {\begin{center}}
\def\ec {\end{center}}
\def\bfg {\begin{figure}}
\def\efg {\end{figure}}
\def\bi {\begin{itemize}}
\def\ei {\end{itemize}}

\def\no {\noindent}

\def\vs {\vspace}

\newcommand{\bdm}{\begin{displaymath}}
\newcommand{\edm}{\end{displaymath}}

\begin{document}

%\preprint{gr-qc/0704.xxxx}
%\hspace{15cm} 03/31/2009\\
\title{On the quantum origin of a small positive cosmological constant
%\footnote{Essay written for the Gravity Research Foundation 2010 Awards for Essays on Gravitation}
}

\author{Saurya Das $^1$} \email[email: ]{saurya.das@uleth.ca}
\author{Rajat K. Bhaduri$^{2}$}\email[email: ]{bhaduri@physics.mcmaster.ca}

\vs{0.3cm}

\affiliation{$^1$
Theoretical Physics Group and
Quantum Alberta,
Department of Physics and Astronomy,
University of Lethbridge, 4401 University Drive,
Lethbridge, Alberta, Canada T1K 3M4 \\}

\affiliation{$^2$ Department of Physics and Astronomy, McMaster University, Hamilton, Ontario,
Canada L8S 4M1 \\}

\begin{abstract}
We show that Dark Matter consisting of ultralight bosons in a Bose-Einstein condensate induces, via its quantum potential,
a small positive cosmological constant which matches the observed value. This explains its origin and why the densities of Dark Matter and Dark Energy are approximately equal.

\end{abstract}
%\pacs{123xxx}

\maketitle

%%%%%%%%%%%%%%%%%%%%%%%%%%%%%%%%%%%%%%%%%%%%%%%%%%%%%%%

It has been established beyond reasonable doubt that the universe is
homogeneous and isotropic at large scales, spatially flat, obeys general relativity and is
made up of $25\%$ cold Dark Matter (DM), $70\%$ Dark Energy (DE) exerting negative presssure
and the rest visible matter \cite{weinberg}. The most favoured candidate for DE is a cosmological constant $\Lambda$.
However the following questions remain unanswered: \\
1. What constitutes DM? \\
2 What constitutes DE/$\Lambda$? \\
3. Why is $\Lambda$ positive? \\
4. Why is $\Lambda$ so tiny, about $10^{-123}\ell_{Pl}^{-2}$
where $\ell_{Pl}$ is the Planck length? \\
5. Why is the current DM and DE contents of universe approximately equal
(the `coincidence problem')?
That is, $\rho_{DM}\approx \rho_{\Lambda} \approx \rho_{crit}$, where
$\rho_{DE}=\rho_\Lambda=\Lambda c^2/8\pi G$,
$\rho_{crit}=3H_0^2/8\pi G
\approx 10^{-26}kg/m^3$ is the critical density, with
$H_0$ the current value of Hubble parameter and $G$ Newton's constant.

We show that if one answers the first question above by assuming that DM is made up of ultralight bosons,
%of mass $m<1~eV/c^2$,
answers to the remaining questions follow.

First as shown in \cite{dasbhaduri1,dasbhaduri2},
the critical temperature of an ideal gas of ultralight bosons of mass $m$,
below which they will drop to their lowest energy state and form a Bose-Einstein condensate (BEC) is given by %
\begin{eqnarray}
T_c = \frac{4.9}{m^{1/3}~ a}~K~,
\end{eqnarray}
where $a$ is the cosmic scale factor (we assume $a=1$ in the current epoch) and $m$ is in $eV/c^2$. In deriving the above, we have equated the DM and BEC densities.
Therefore for $m<6~eV/c^2$,
$T(a)< T_c(a)~\forall a$, where $T(a)=2.7/a$, the ambient temperature of the universe. A BEC of these ultralight bosons will therefore form in the early universe
\footnote{For a list of earlier papers on BEC in cosmology, see \cite{dasbhaduri1,dasbhaduri2}.}.
We assume the average BEC density to be
$\rho_{crit}$ at present.
%
%
%
%The BEC will be described by a
%macroscopic wave function
%$\Psi={\cal R} e^{iS/\hbar}$,
%where ${\cal R}, S \in \mathbb{R}$.
%To determine the wave function, consider
%
Although the bosons start-off as ultrarelativistic particles, once inside the BEC, they are slow and non-relativistic. This and the facts that the spatial curvature of the universe is zero and its spacetime curvature is negligible
(also $\approx 10^{-123}\ell_{Pl}^{-2}$) justifiy the use of Newtonian gravity. For example, a particle of mass $m$ in the condensate anywhere within the homogeneous and isotropic universe, and on the surface of a uniform sphere of radius $r$ with density $\rho_{crit}$, is subjected to gravity from matter $M$ contained within that sphere (matter outside the sphere does not count) obeying the equation
\begin{equation}
m \ddot{r}=-\frac{GMm}{r^2}.
\label{ho1}
\end{equation}
Using $M=\frac{4 \pi}{3} \rho_{crit} r^3$, the above becomes just the equation of a harmonic oscillator with
angular frequency given by $\omega^2 = 4\pi G\rho_{crit}/3=H_0^2/2
=c^2/(2L_0^2)$,
where $L_0$ is the Hubble radius. Plugging in $r=r_0~a(t)$,
where $r_0$ a constant,
we get the Friedmann equation with $\rho=\rho_{crit}$
\begin{eqnarray}
\frac{\ddot a}{a}
= - \frac{4\pi G}{3}\rho_{crit}
= -  \omega^2 .
\label{ho2}
\end{eqnarray}
Therefore with just DM, one would end up with an oscillating scale factor.
%and fluctuating universe.
%
We now show that quantization changes things entirely.
Since the bosons in the BEC all have the same wave function and are nonrelativistic, we can use the one-particle Schr\"odinger equation
\begin{equation}
-\frac{\hbar^2}{2m} \nabla^2 \Psi + V \Psi = i\hbar\frac{\partial \Psi}{\partial t}.
\label{cos2}
\end{equation}
with $V=m\omega^2 r^2/2$.
Writing the BEC wave function
$\Psi$ as
\begin{equation}
\Psi (\vec r,t)={\cal R} %\exp
e^{i\frac{S}{\hbar}}~,
\label{psi1}
\end{equation}
where ${\cal R} (\vec r,t)$ and $S (\vec r,t)$ are real functions.
Substituting
Eq.(\ref{psi1}) in Eq.(\ref{cos2}),
one obtains the following equations
\cite{bohm}
%
%\begin{equation}
%-\frac{\partial S}{\partial t}=\frac{(\nabla S)^2}{2m}+ V(r)
%-\frac{\hbar^2}{2m}
%\frac{\nabla^2 {\cal R}}{{\cal R}}~,
%\label{cos3}
%\end{equation}
%
%and the continuity equation
%
\begin{eqnarray}
&& \frac{\partial \rho}{\partial t}+\nabla.(\rho \vec v)=0,
\label{cont1} \\
&& m \frac{d\vec v}{dt}
= -\vec\nabla V
+ \frac{\hbar^2}{2m}\vec \nabla
\left( \frac{1}{\cal R} \nabla^2 {\cal R}
\right) \label{nl1}
\end{eqnarray}
with $\rho=|\Psi|^2={\cal R}^2$
and $\vec v = \vec \nabla S/m$.
Eq.(\ref{cont1}) is the continuity equation for the probability current,
$\vec J =\rho\vec v$
while Eq.(\ref{nl1})
represents Newton's law with the additional `quantum potential' term,
$V_Q=-\frac{\hbar^2}{2 m}
\frac{\nabla^2 {\cal R}}{{\cal R}}$,
which depends on the wave function of the system and vanishes in the classical or
$\hbar\rightarrow 0$ limit.
For the macroscopic BEC wave function, one chooses the harmonic oscillator ground state,
since the vast majority of bosons in the BEC will be in that state. That is,
\begin{eqnarray}
\Psi = R(a)~e^{-r^2/\sigma^2}~,
\label{psi2}
\end{eqnarray}
where $\sigma^2=2\hbar/m\omega= 2\sqrt[]{2}\lambda L_0$,
$\lambda=\hbar/mc$ is the Compton wavelength of
the constituent bosons.
We have omitted the time-dependent phase factor, since it plays no role in the quantum potential.
The scale factor has an intrinsic time-dependence, which we assume to be slowly varying. The dilution of DM with time is then entirely accounted for
by choosing
$R(a)=R_0/[a(t)]^{3/2}$, such that
$\rho_{crit}\propto R_0^2$,
DM density $\rho_{DM} \propto |{\cal R}|^2 \propto 1/a^3$
and BEC particle number is conserved in time.
It can be shown that $\sigma$
remains approximately a constant
over a time-period of oscillation.
Straightforward calculation then
gives for the wave function in
Eq.(\ref{psi1}),
\begin{eqnarray}
V_Q &&
%=\frac{\hbar^2}{2m \sigma^2} \left(6-\frac{4r^2}{\sigma^2}\right)
= \frac{3}{2}~\hbar\omega - \frac{1}{2} m\omega^2 r^2 \label{vqho} \\
&& = \frac{3\hbar^2}{m\sigma^2} - \frac{2\hbar^2 r^2}{m\sigma^4}~.
\end{eqnarray}
Substituting $-\vec\nabla V_Q/m$ in the RHS of Eq.(\ref{ho1}), one now gets
the quantum corrected Friedmann equation
\begin{eqnarray}
\frac{\ddot a}{a} &&= -\frac{4\pi G \rho_{crit}}{3}
+ \frac{\Lambda c^2}{3} \\
&& =-\frac{4\pi G}{3}
\left( \rho_{DM} + \rho_{\Lambda}
+ 3p_{\Lambda} \right) = 0~,
\end{eqnarray}
where
%the quantum potential induced cosmological constant
$\Lambda=3H_0^2/2c^2=3/(2L_0^2)$.
Therefore
$\rho_{\Lambda}=
\Lambda c^2/8\pi G
=\rho_{crit}/2$ and
$p_{\Lambda}=
-\Lambda c^2/8\pi G
=-\rho_{\Lambda}$.
In other words, the quantum potential induces a positive cosmological constant
$\Lambda$,
whose density (approximately) equals the DM density, has negative pressure, and the Einstein static universe is recovered at
$a=1$!
The required small positive $\Lambda$ comes out automatically due to quantum mechanics and there is no need to put it in by hand.
$\rho_{DM}$ continues to decay as the
universe evolves, resulting in
$\rho_{DM}<\rho_{\Lambda}$
for $a>1$.

In summary, in this letter we have
shown that the quantum potential of the coherent DM BEC wave function
induces a cosmological constant
with density equal to DM density.
This solves the coincidence problem.
Furthermore, $\Lambda$ is small because $\rho_{DM}\approx \rho_{crit}$ is small in the current epoch. We have also
provided a simple explanation as to why $\Lambda$ is positive - because the quantum potential, a manifestation of the uncertainty principle, is intrinsically positive. Note that the preferred
boson mass $m \approx 10^{-22}~eV/c^2$, which prevents the formation of small-scale structure in DM also via the uncertainty principle \cite{hu,marsh},
is well within the allowed mass range for BEC.
In other words, this simple model of BEC DM answers all the questions posed at the beginning. The macroscopic wave function is essential for it to work.
The slight excess of
$\rho_{\Lambda}$ over
$\rho_{DM}$ %=0.25\rho_{crit}$
%=0.70\rho_{crit}$
in reality and the resulting accelerating universe
can be attributed to the continuous dilution of DM,
$\omega$ not being strictly a constant, that not 100\% of the bosons are in the ground state
and that we have ignored self-interaction of the bosons.
As mentioned earlier, use of Newtonian cosmology is perfectly valid, although arriving at the same results using general relativity is straightforward \cite{sd,dasessay}.
We assumed a slowly varying $\rho_{crit}$, which is guaranteed soon after the quantum potential is generated, since
$\rho=\rho_{DM} + \rho_{DE}$ and although the former decays as $1/a^3$, the latter remains constant in time.
Note that our result should be valid for any epoch. This means that there will be a decrease in the value of $\Lambda$ over time, possibly in discrete jumps. This and the consequent slight increase in the estimated age of the universe should have observable consequences.
Our model does not explain why our universe is spatially flat, but accepts it as an observational fact.
And while we give the allowed range of boson mass, we do not speculate on what these bosons are.
Nor do we speculate on the evolution of the universe when the DM wave function gets flat enough for its quantum potential to decay.
More work needs to be done to answer these questions.

%Our model also suggests the following evolution of the universe:
%at very late times, the DM density and the related wave function will be diluted and $V_Q$ would decay to zero. The universe would start to collapse resulting in increased density and when sufficient $\rho_{DM}$ is accumulated, it will
%again induce a $\rho_\Lambda$
%leading to expansion, and the cycle continues.
%

%\vs{.2cm}
%%%%%%%%%%%%%%%%%%%%%%%%%%%%%%%%%%%%%%%%%%%%%%%%%%%%%%
\no {\bf Acknowledgment}

\no
This work is supported by the Natural Sciences and Engineering
Research Council of Canada.
%

%%%%%%%%%%%%%%%%%%%%%%%%%%%%%%%%%%%%%%%%%%%%%%%%%%%%%

%\section*{References}

\noindent
%\underline{Author Contribution Statement}
%Saurya Das and Rajat K. Bhaduri contributed equally to the idea and writing of the paper.

\vspace{0.5cm}
\noindent
%\underline{Competing financial interests:}
%there are no competing financial interests

\noindent
%\underline{Competing non-financial interests:}
%there are no competing non-financial interests

\end{document}